\documentclass[
    ,final            
  ]
  {aipproc}

\layoutstyle{6x9}

\begin{document}

\title{Gravity from Lorentz Symmetry Violation}

\classification{04.20.Cv, 11.30.Cp, 11.15.Ex, 95.30.Sf}
\keywords      {Lorentz symmetry, spontaneous symmetry breaking, gravitation}


\author{Robertus Potting}{
  address={CENTRA and Physics Department, FCT, University of the Algarve,
 8005-139 Faro, Portugal}
}

\begin{abstract}
In general relativity, the masslessness of gravitons can be traced to
symmetry under diffeomorphisms. Here we consider another possibility,
whereby the masslessness arises from spontaneous violation of Lorentz
symmetry.
\end{abstract}

\maketitle


In the standard approach to general relativity, the Einstein
equations are derived using geometrical notions of Riemannian space-time
such as curvature. Another approach exists \cite{otherapproach}, in which the starting point
is the linearized free equation of motion for a spin-2 particle
(the graviton)
\begin{equation}
R^L_{\mu\nu}\equiv K_{\mu\nu\alpha\beta}h^{\alpha\beta}=0
\label{eq:free_einstein}
\end{equation}
which can be obtained from a free lagrangian $\mathcal{L}^0$.
From Newton's law we know gravity is coupled to matter. The way to
take this into account consistently is to add a term proportional
to the matter energy-momentum tensor.
In the case of free gravitons described by (\ref{eq:free_einstein}) we
have no matter, and the equation is consistent as it stands. Nevertheless,
gravitons represent energy-momentum, and thus should be expected to
contribute as well to the right-hand-side of (\ref{eq:free_einstein}).
The resulting, modified equation of motion
can be obtained from a cubic lagrangian $\mathcal{L}^1$.
In turn, the cubic term in this lagrangian gives another contribution
to the energy-momentum tensor, which should again be added to the
right-hand-side of (\ref{eq:free_einstein}). This leads to a quartic
lagrangian $\mathcal{L}^2$. This process continues indefinitely and in the
limit one recovers an equation that can be summarized by the full Einstein
equation in free space
\begin{equation}
R_{\mu\nu}=0.
\label{eq:full_einstein}
\end{equation}
A convenient one-step algorithm that leads from (\ref{eq:free_einstein})
to (\ref{eq:full_einstein}) has been derived by Deser \cite{Deser}.

In this derivation, the reason for starting with a symmetric field
$h^{\mu\nu}$ is easily understood: it is needed in the action to
couple to the symmetric energy-momentum tensor $T^{\mu\nu}$.
What is the reason for the masslessness of gravitons?

The usual argument that is given is one of symmetry. For instance,
in the electroweak model, the photon is massless because of an unbroken
$U(1)$ gauge symmetry. In QCD, the gluons are massless because
of the $SU(3)$ gauge symmetry. Similarly, in general relativity
the masslessness of the graviton is taken to be a consequence of
diffeomorphism symmetry.

However, an alternative reason exists as to why a particle might be
massless, which has to do with the breaking of a symmetry, rather than
its presence. The Nambu-Goldstone theorem states that, with some mild
assumptions, there must be a massless particle whenever a continuous
global symmetry of an action isn't a symmetry of the vacuum
\cite{NambuGoldstone}. This result is readily understood by
considering an action with a scalar potential $V$ which has its
minimum for nonzero field values. Considering a vacuum in which the
field assumes such a constant value, a symmetry of the theory shifts
the vacuum to another, equivalent, minimum. Thus $V$ has at least one
flat direction; excitations around the vacuum in that direction
correspond to massless particles, the Nambu-Goldstone modes.

A well-known example of the realization of this mechanism is the
(approximate) masslessness of the pion due to the spontaneous breaking
of chiral symmetry in the sigma model. A more recent example involves
the so-called ``bumblebee'' model \cite{bumblebee,BluhmKostelecky} with
Lagrangian
\begin{equation}
\mathcal{L}_B=-{1\over4}B_{\mu\nu}B^{\mu\nu}-\lambda(B_\mu B^\mu\pm b^2)
-B_\mu J^\mu.
\label{eq:bumblebee}
\end{equation}
Here $B_{\mu\nu}=\partial_\mu B_\nu-\partial_\nu B_\mu$, $\lambda$
is a Lagrange multiplier field and $J^\mu$ is an external current.
The equation of motion of the latter
forces the vector $B_\mu$ to assume a nonzero vacuum value $b_\mu$,
breaking spontaneously the Lorentz symmetry. Thus we expect massless
Goldstone modes in the flat direction of the potential, which can be
obtained by Lorentz transformations of the vacuum value:
\begin{equation}
\delta B^\mu\equiv A^\mu=B^\mu-b^\mu \approx \epsilon^{\mu\nu}b_\nu.
\end{equation}
Here the $\epsilon^{\mu\nu}$ generate a Lorentz transformation.
Expressing (\ref{eq:bumblebee}) in terms of $A_\mu$ yields
\begin{equation}
\mathcal{L}_B\to\mathcal{L}_{NG}\approx-{1\over4}F_{\mu\nu}F^{\mu\nu}
-A_\mu J^\mu-b_\mu J^\mu
\end{equation}
subject to the constraint $b_\mu A^\mu=0$. Thus we obtain electrodynamics
in the axial gauge! Note that, in this model, the masslessness of the
photon arises, not by the presence of a $U(1)$ gauge symmetry (the
Lagrangian (\ref{eq:bumblebee}) is not gauge invariant), but by
the spontaneous breaking of Lorentz symmetry.

Instead of the Lagrange multiplier term in (\ref{eq:bumblebee}) one can
take a smooth scalar potential with a nonzero minimum. In that case
we have, apart from the Nambu-Goldstone fluctuations in the flat direction,
radial fluctuations corresponding to a massive particle.

As it turns out, spontaneous breaking of Lorentz symmetry can be
employed as well to obtain a massless graviton \cite{Kostelecky:2005ic}.
To this effect, we consider the Lagrangian density
\begin{equation}
\mathcal{L}_C={1\over2}C^{\mu\nu}K_{\mu\nu\alpha\beta}C^{\alpha\beta}
-V(C^{\mu\nu}).
\label{eq:cardinal}
\end{equation}
Here, $C^{\mu\nu}$ is a symmetric two-tensor, $K_{\mu\nu\alpha\beta}$
is the usual quadratic kinetic operator for a massless spin-2 field,
and $V(C^{\mu\nu})$ is a scalar potential constructed from
$C^{\mu\nu}$ and $\eta_{\mu\nu}$. This theory is invariant under local
Lorentz transformations and under diffeomorphisms. The potential $V$
ensures $C^{\mu\nu}$ takes some vacuum value $c^{\mu\nu}$, which we
will assume to be nonzero. This spontaneously breaks local Lorentz and
diffeomorphism invariances. The massless Nambu-Goldstone fields are
the excitations $\delta C_{\mu\nu}= C_{\mu\nu}-c_{\mu\nu}$ about this
solution, generated by the broken symmetries and maintaining the
potential minimum:
\begin{equation}
\delta C_{\mu\nu}= C_{\mu\nu}-c_{\mu\nu}\approx
\epsilon_\mu{}^\alpha c_{\alpha\nu}+\epsilon_\nu{}^\alpha c_{\alpha\mu}
\equiv M_{\mu\nu}{}^{\alpha\beta}\epsilon_{\alpha\beta}
\end{equation}
with $M_{\mu\nu\alpha\beta}={1\over2}(\eta_{\mu\alpha}c_{\nu\beta}+
\eta_{\nu\alpha}c_{\mu\beta}-\eta_{\mu\beta}c_{\nu\alpha}-
\eta_{\nu\beta}c_{\mu\alpha})$. For generic $c_{\mu\nu}$, these
fluctuations satisfy the four (linearized) constraints
\begin{equation}
\delta C^\mu{}_\nu(c^m)^\nu{}_\mu=0
\end{equation}
with $m=0,1,2,3$.

The equations of motion that follow from varying the lagrangian
density (\ref{eq:cardinal}) with respect to the independent degrees of
freedom $\epsilon_{\mu\nu}$ are $M^{\mu\nu\rho\sigma}K_{\mu\nu\alpha\beta}
\delta C^{\alpha\beta}=0$, that can be solved using Fourier decomposition.
The solutions obey the massless wave equation
\begin{equation}
\partial^\lambda\partial_\lambda \delta C_{\mu\nu}=0,
\end{equation}
subject to the Lorenz condition $\partial^\mu\delta C_{\mu\nu}=0$.
The Lorenz condition fixes four of the initial six independent degrees
of freedom carried by the Lorentz generators, leaving two massless
propagating degrees of freedom.

It is not difficult to show that this theory matches the usual
description of a massless graviton field $h^{\mu\nu}$ in Minkowski
spacetime. Consider the Lagrange density for a free
massless graviton $h^{\mu\nu}$:
\begin{equation}
\mathcal{L}_h={1\over2}h^{\mu\nu}K_{\mu\nu\alpha\beta}h^{\alpha\beta}.
\label{eq:linear_gravity}
\end{equation}
Initially $h^{\mu\nu}$ has ten degrees of freedom. The Lagrangian
(\ref{eq:linear_gravity}) is invariant under local diffeomorphisms
\begin{equation}
\delta h_{\mu\nu}=\partial_\mu\xi_\nu+\partial_\nu\xi_\mu.
\end{equation}
Consequently, we can choose four gauge fixing conditions. Rather than
adopting the usual transverse-traceless gauge, we pick a different gauge
that yields a direct match with the new theory. For generic $c_{\mu\nu}$,
we choose the conditions
\begin{equation}
h^\mu{}_\nu(c^m)^\nu{}_\mu=0
\end{equation}
with $m=0,1,2,3$.  From the equations of motion
$K_{\mu\nu\alpha\beta}h^{\alpha\beta}=0$ and the gauge conditions one
finds that the solutions satisfy the usual wave equation for a
massless field $\partial^\lambda\partial_\lambda h_{\mu\nu}=0$, as
well as the Lorenz condition $\partial^\mu h_{\mu\nu}=0$.  This leaves
$10-4-4=2$ propagating degrees of freedom, and an explicit match with
the theory of the cardinal field. Once more, we see that while the
symmetry structure of the two theories is radically different, the
equations are in direct correspondence at low energy.

The full nonlinear Einstein equations can be obtained by insisting on
a consistent coupling to the energy-momentum tensor, by using a
version of Deser's one-step procedure referred to above.

While reproducing the Einstein equations at lowest order, the new
theory differs from general relativity in various ways.  

In the pure-gravity sector, the new theory has subleading corrections
to the Einstein equations in vacuum. They are of higher order in the
Riemann tensor, and thus very small for nearly Minkowski spacetimes
(and thus in laboratory and solar-system tests). However, they may
lead to significant deviations from general relativity in extreme
environments such as black holes, or the early Universe.

There are effects in the matter sector because of couplings of the
type $c_{\mu\nu}T^{\mu\nu}$ involving the matter energy-momentum
tensor.  A framework for the comprehensive treatment of such effects
exists that maintains standards of consistency such as stability and
microcausality \cite{SME}.  Numerous experimental searches looking for
Lorentz-violating signals within this framework are currently under
way.

Most interestingly, maybe, are structural differences related to
excitations of the cardinal field in the non-flat directions. These
will play a role at very high energies (presumably close to the Planck
scale) and temperatures. Also, at high temperatures the potential $V$
acquires corrections that restore local Lorentz symmetry by shifting
the minimum of the effective potential to zero cardinal field value.
These effects will have profound implications for the very early
Universe.

It is clear that the quantum properties of the new model will differ
significantly from those of general relativity. It the case of the
bumblebee model it has been shown that nonpolynomial and superficially
unrenormalizable potentials $V$ can become renormalizable and stable
when quantum corrections are included \cite{Altschul:2005mu},
a result that can be extended to the current gravity theory.

\begin{theacknowledgments}
This talk is based on work done in collaboration with V.\ Alan
Kosteleck\'y.  R.P.\ acknowledges financial support from the Funda\c
c\~ao para a Ci\^encia e a Tecnologia (Portugal), and wishes to thank
the Physics Department of Indiana University for hospitality.
\end{theacknowledgments}

\end{document}